\documentclass[fleqn,twoside]{article}
\usepackage{espcrc2}
\input{epsfig.sty}
\def\Journal#1#2#3#4{{#1} {\bf #2}, #3 (#4)}

\def\NPA{{\rm Nucl. Phys.} A}

\def\PLB{{\rm Phys. Lett.}  B}
\def\PRL{\rm Phys. Rev. Lett.}

\def\PRD{{\rm Phys. Rev.} D}

\def\la{\langle}
\def\ra{\rangle}
\def\be{\begin{equation}}
\def\ee{\end{equation}}
\def\bea{\begin{eqnarray}}
\def\eea{\end{eqnarray}}
\def\lsim{\mathrel{\rlap{\lower4pt\hbox{\hskip1pt$\sim$}}
    \raise1pt\hbox{$<$}}}         
\def\gsim{\mathrel{\rlap{\lower4pt\hbox{\hskip1pt$\sim$}}
    \raise1pt\hbox{$>$}}}

\usepackage{graphicx}
\usepackage[figuresright]{rotating}

\newcommand{\AmS}{{\protect\the\textfont2
  A\kern-.1667em\lower.5ex\hbox{M}\kern-.125emS}}

\title{Light-Front QCD Hamiltonian Dynamics and Constituent Quark Picture
in Exclusive Processes}

\author{Chueng-Ryong Ji\address{Department of Physics, North Carolina State
                         University,\\
        Raleigh, NC 27695-8202, U.S.A}%
        \thanks{ji@ncsu.edu}
        and
        Ho-Meoyng Choi\address[MCSD]{Department of Physics,
        Carnegie Mellon University, \\
        Pittsburgh, PA 15213, U.S.A}%
        \thanks{homeoyng@andrew.cmu.edu}
        }
\begin{document}

\begin{abstract}
After reviewing a connection between quantum chromodynamics and contituent
quark model pictures in the light-front quantization with some
comparison and contrast to the
ordinary equal-time bridge a la Bogoliubov-Valatin transformation,
we discuss some newer development of the light-front quark model
phenomenology in exclusive processes including the embedded state. The
skewed parton distribution appears to
be a good testing ground for our new effective treatment of the light-front
nonvalence contributions in timelike exclusive processes.
\vspace{1pc}
\end{abstract}

\maketitle


One of the most puzzling features in hadron physics is the connection
between the two fundamentally different pictures of hadronic matter, i.e.
the quantum chromodynamics (QCD) based on a covariant non-Abelian
quantum field theory and the constituent quark model (CQM) closely
related to experimental observations.
While the QCD has a complicate nonperturbative vacuum structure
manifested by the color confinement and the dynamical breaking of the
chiral symmetry, the CQM is mostly built on a rather simple vacuum.
Because of the bound state problem inherent in the hadron physics, the
Hamiltonian approach based on an equal-time formulation is often used to
remove the complications from the relative time degrees of freedom among
the constituents. In this respect, both the ordinary time $t$ and the
light-front(LF) time $\tau = t+z/c$ are by far the most popular choices for
an equal-time formulation. However, these two choices render quite different
pictures for the bridge between QCD and CQM.

In the equal-$t$ approach, one needs to transform a complicate QCD
vacuum to a rather simple CQM vacuum. To do this, people have
often utilized the Bardeen-Cooper-Schrieffer(BCS) type
Bogoliubov-Valatin (BV) transformation~\cite{SSJC,RSSJC,GJC1}.
With this transformation, one can
get a coherent vacuum for quasiparticle constituents and the mass gap
equation providing a relation beween the current quark mass and the
consituent quark mass. The QCD Hamiltonian $H_{QCD}$ in the Coulomb gauge
can be split into $H_0$ describing the ball park of the physical system
and $H_I$ being the rest of the Hamiltonian that can give small
perturbative corrections. This splitting can be done by adding and
subtracting a phenomenological Hamiltonian $H_{phen}$ to the kinetic
energy part $K$ and the interaction part $H^I_{QCD}$, respectively, i.e.
\begin{eqnarray}
H_{QCD} &=& (K + H_{phen}) + (H^I_{QCD} - H_{phen}) \\ \nonumber
&=& H_0 + H_I.
\label{hamiltonian}
\end{eqnarray}
While $H_{phen}$ is often given by the confining (e.g. linear) potental,
the residual interaction $H_I$ is ususally taken as the canonical QCD
interaction Hamiltonian at the cutoff scale $\Lambda $ infinite;
$H_I \approx H^I_{can}(\Lambda\rightarrow\infty)$.
However, in order to find a low energy effective
Hamiltonian, one needs to set $\Lambda$ finite and lower it to the scale
of the effective Hamiltonian that one wants to find. Introducing a finite
$\Lambda$ breaks the symmetry of the orginal Hamiltonian and to recover it
one needs a counter term $H_{CT}(\Lambda)$. Lowering $\Lambda$ generates
the part of effective Hamiltonian $H_{gen}(\Lambda)$ that compensates the
physics between the two $\Lambda$ values before and after lowering it.
Both $H_{CT}(\Lambda)$ and $H_{gen}(\Lambda)$ can be found by either
a similarity renormalization procedure~\cite{RSSJC}
or a flow equation method~\cite{GJC1}.

In the equal-$\tau$ (LF) approach~\cite{BPP}, however, the
procedure of
connecting between QCD and CQM may be drastically different from
the equal-$t$ case since the vacuum at equal $\tau$ has a dramatic difference
compare to the vacuum at equal $t$. For the particle which has
the mass $m$ and the four-momentum $k = (k^{0},k^{1},k^{2},k^{3})$,
the relativistic energy-momentum relation at equal $\tau$ is
given by
\begin{equation}
k^{-} = \frac{{{\bf k}^{2}_{\perp i}} + m^{2}}{k^{+}},
\label{lfdr}
\end{equation}
where the LF energy conjugate to $\tau$ is given by $k^{-} = k^{0}
- k^{3}$ and the LF momenta $k^{+} = k^{0} + k^{3}$ and
${{\bf k}_{\perp}} = (k^{1},k^{2})$ are orthogonal to $k^{-}$ and form
the LF three-momentum $\underline{k} = (k^{+},{{\bf k}_{\perp}})$.
The rational relation given by Eq.(\ref{lfdr})
provides a remarkable feature to the LF vacuum, namely , the
Fock state vacuum is an eigenstate of the full Hamiltonian.
Consequently, all bare quanta in
hadronic Fock states are associated
with the hadron and none are disconnected elements of the vacuum.
This leads to a relatively simple vacuum structure in QCD. There is no
spontaneous creation of massive fermions in the LF quantized vacuum.
Thus, one can immediately obtain a constituent-type picture, in which
all partons in a hadronic state are connected
directly to the hadron
instead of
being simply disconnected excitations (or vacuum fluctuations) in a
complicated medium. Nevertheless, one needs to cutoff 
the zero-modes~\cite{zero}
corresponding to the degrees of freedom with $k^+ \rightarrow 0$ to obtain a
salient CQM picture because the zero-mode fluctuations are still possible
in the LF vacuum. The cutoff of the zero-modes introduces
the mass scale of the constituent quark as well as the counter term that
can generate a non zero amplitude of particle creation and
is therefore a
possible source for the features associated with a nontrivial vacuum
structure including confinement and spontaneous symmetry breaking.

Recently, the flow equation method of continuous unitary transformations
was used to eliminate the minimal quark-gluon interaction in the LF
$H_{QCD}$~\cite{GJC2}. Dividing the complete Fock space into two components,
a tractable $P$ subspace spanned by
states with a small number of quanta and the remainder $Q =1 - P$, one can
get a Hamiltonian matrix to be diagonalized of the form
\begin{eqnarray}
   H = \left(
   \begin{array}{cc}
      PHP & PHQ \\
      QHP & QHQ
   \end{array}\right)
\,.\label{block}
\end{eqnarray}
The flow equation method leads to block diagonalize this matrix
yielding the effective Hamiltonian

\begin{eqnarray}
   H_{ eff} =\left(
   \begin{array}{cc}
     PH_{ eff}P & 0            \\
     0             & QH_{ eff}Q
   \end{array}\right)\,.
\end{eqnarray}
One can then separately diagonalize the two blocks which are now
uncoupled. The coupled differential equations in the two lowest Fock
sectors correspond to the renormalization of the LF gluon mass
and the generation of an effective quark-antiquark (as well as
gluon-gluon) interaction. From these, a more singular $1/q^4$ behavior
for the quark and gluon effective interactions at small gluon momenta can
be obtained~\cite{GJC2}. Therefore, one can catch a glimpse of support
from the LF QCD to
the LF CQM which we call LF quark model
(LFQM)~\cite{CJ}.

Indeed, there has been a significant progress in describing the meson
properties with the LFQM for the spacelike region~\cite{CKJ}.
Again, the success of this model hinges upon the simplicity of LF vacuum.
The complicated nontrivial vacuum effect from the zero-modes has been 
traded off by the constituent quark masses. Moreover, the Drell-Yan-West 
($q^+ = q^0 + q^3 = 0$) frame in the LF
quantization provided an effective formulation for the calculation of
various form factors in the spacelike momentum transfer region $q^2 =
-Q^2 < 0$~\cite{LB}.
As an example, only the valence diagram shown in Fig. 1(a) is needed in
$q^+ = 0$ frame when the ``good" components of the
current, $j^{+}$ and $j_{\perp}=(j_{x},j_{y})$, are used
for the spacelike electromagnetic form factor calculation of pseudoscalar
mesons. Successful LFQM description of various hadron form
factors can be found in the literatures~\cite{CJ,Chung}.
\begin{figure}
\psfig{figure=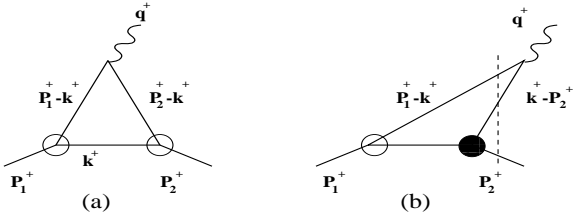,width=3in,height=1.1in}
\vspace{-1.3cm}
\caption{(a) The usual light-front valence diagram and (b) the
nonvalence(pair-creation) diagram. The vertical dashed line in (b)
indicates the energy-denominator for the nonvalence contributions.}
\end{figure}

However, not all is well.
The timelike ($q^{2}>0$) form factor analysis in the LFQM
has been hindered by the fact that $q^{+}=0$ frame is defined
only in the spacelike region ($q^{2}=q^{+}q^{-}-q^{2}_{\perp}<0$).
While the $q^{+}\neq0$ frame can be used in principle to compute the timelike
form factors, it is inevitable (if $q^{+}\neq 0$) to encounter the nonvalence
diagram arising from the quark-antiquark pair creation (so called
``Z-graph"). For example, the nonvalence diagram in the case of semileptonic
meson decays is shown in Fig. 1(b).
The main source of the difficulty, however, in calculating the nonvalence
diagram (see Fig. 1(b)) is the lack of information on the embedded
state represented by the black
blob which should contrast with the white blob representing the usual LF
valence wave function.

\begin{figure}
\psfig{figure=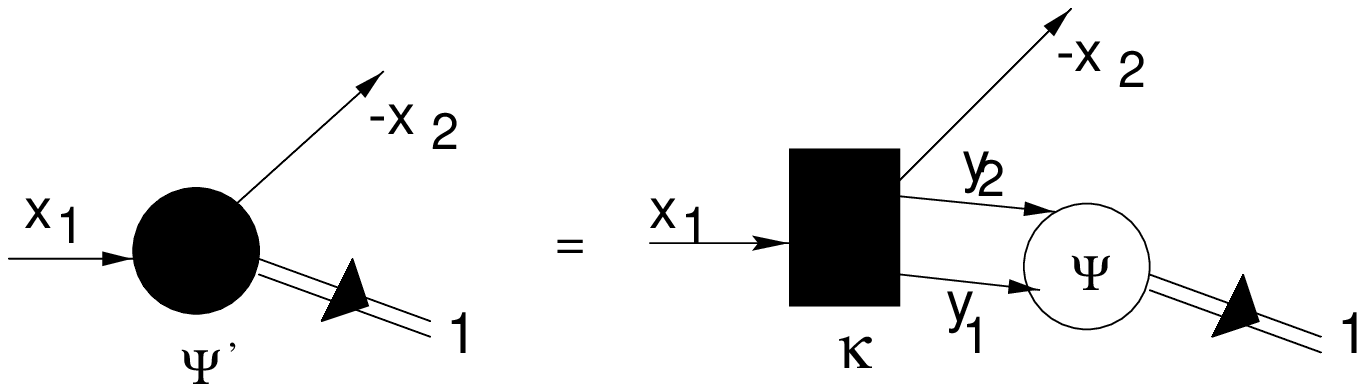,height=0.7in,width=2.5in}
\vspace{-0.8cm}
\caption{Non-wave-function vertex(black blob) linked to an ordinary
LF wave function(white blob).\label{SD3}}
\end{figure}

Fortunately, we've recently came up with an effective way of handling the
nonvalence contribution~\cite{JC}. Our aim of new treatment was to make the
program suitable for the CQM phenomenology specific to the low momentum
transfer processes.
The key of our method is the link between the non-wave-function vertex
(black blob) and the
ordinary LF wave function (white blob) as shown in Fig.~\ref{SD3}, i.e.,
\bea\label{eq:SD}
&&\hspace{-0.8cm}(M^2-M'^{2}_0)\Psi'(x_i,{\bf k}_{\perp i})\nonumber\\
&=&\hspace{-0.2cm}\int[dy][d^2{\bf l}_{\perp}]
K(x_i,{\bf k}_{\perp i};y_j,{\bf l}_{\perp j})
\Psi(y_j,{\bf l}_{\perp j}),
\eea
where $M$ is the mass of outgoing meson and $M'^{2}_0=(m^2_1+{\bf
k}^2_{\perp 1})/x_1 - (m^2_2+{\bf k}^2_{\perp 2})/(-x_2)$ with
$x_1 = 1-x_2 > 1$ due to the kinematics of the non-wave-function vertex.

To discuss further details of our method~\cite{JC}, we
formulate the pion form factor in terms of the
off-forward parton distribution functions, the so called ``skewed parton
distributions (SPDs)" that are the generalization of the ordinary
(forward) distribution functions~\cite{CJK}.
A well-known and practical example of SPDs as a nonperturbative
information entering the LF dominated hard scattering processes
is the deeply virtual Compton scattering (DVCS) $\gamma^* p\to\gamma p$ for
large initial photon virtuality $Q^2$ and small $t$ region,
which can be factorized into a hard photon-parton and a skewed parton
distribution~\cite{Mu,XJ1,Ra1}.
Since the usual local photon vertex in
the pion form factor analysis is replaced by a nonlocal operator of the
SPDs, one can explore new physics.

In the LF coordinates, the SPDs are in general functions
of the longitudinal momentum fraction variable $x$, the skewedness parameter
$\xi=(P-P')^+/P^+$ measuring asymmetry between initial ($P$) and
final ($P'$) hadron state momenta,
and the squared momentum transfer $t$.
Analogous to the pion electromagnetic (EM) form factor calculation
\be\label{EM}
J^+(0)\equiv\langle P'|{\bar\psi(0)}\gamma^+\psi(0)|P \rangle
=F_{\pi}(t)(P+P')^+,
\ee
we define the SPD
${\cal F}_{\pi}(\xi,x,t)$ of a pion by
\begin{eqnarray}\label{SPD}
{\cal J}^+&\equiv&\int\frac{dz^-}{4\pi}e^{i x P^+z^-/2}
\la P'|{\bar\psi}(0)\gamma^+\psi(z)|P\ra,
\nonumber \\
&=&{\cal F}_{\pi}(\xi,x,t)(P+P')^+,
\end{eqnarray}
where  $z=(z^+,z^-,{\bf z}_\perp)$ in a LF representation
and $z^+={\bf z}_{\perp}=0$.
The SPDs display characteristics of
the ordinary(forward) quark distribution in the limit of $\xi\to 0$
and $t\to 0$, on the other hand,  the first moment of the SPDs
is related to the form factor by the following sum
rules~\cite{XJ1,Ra1}:
\be\label{sum}
\int^1_{0}dx\; {\cal F}_{\pi}(\xi, x, t) = F_\pi(t),
\ee
where ${\cal F}_{\pi}(\xi, x, t)=e_u{\cal F}^u_{\pi}(\xi, x, t)
-e_d{\cal F}^{\bar d}_{\pi}(\xi, x, t)$ and we assume
isospin symmetry($m_u=m_{\bar d}$) so that
${\cal F}^u_{\pi}(\xi, x, t)={\cal F}^{\bar d}_{\pi}(\xi, x, t)$.
Note that Eq.~(\ref{sum}) is independent of $\xi$, which provides
important constraints on any model calculation of the SPDs.
\begin{figure}
\centerline{\psfig{figure=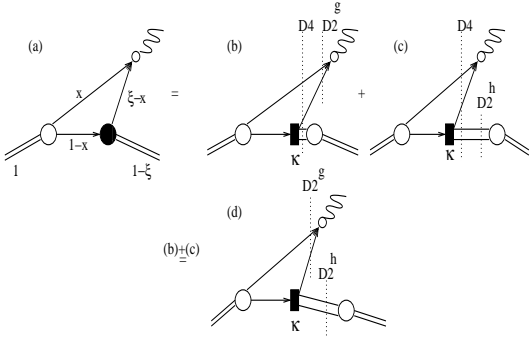,height=4.5cm,width=7cm}}
\vspace{-0.8cm}
\caption{Effective treatment of the LF nonvalence amplitude.
\label{Embed}}
\end{figure}

Considering the quark-meson and quark-gauge boson vertices together,
we also find that the four-body energy denominator ($D_4$)
appearing in Fig.~\ref{Embed} is absent
due to the sum of two possible diagrams
in the LF time-ordering (see Figs.~\ref{Embed}(b) and~(c)).
Summing over the two time-ordered
diagrams Figs.~\ref{Embed}(b) and~(c), one can easily find the following
identity, $1/D_4D^g_2+1/D_4D^h_2=1/D^g_2D^h_2$, which removes the complicate
four-body energy denominator term. We thus obtain the amplitude corresponding
to the nonvalence contribution in terms of
ordinary LF wave functions of hadron and
gauge boson as shown in Fig.~\ref{Embed}(d).
This method, however, requires the relevant
operator ${\cal K}(x,{\bf k}_\perp; y,{\bf l}_\perp)$ which is in general
dependent on the involved momenta connecting
the one-body to three-body sector as depicted in Fig.~\ref{SD3}.
The details of the valence and nonvalence
contributions to the SPDs of the pion in LFQM can be found
in Ref.~\cite{CJK}.
While the relevant operator ${\cal K}$
is in general dependent on all internal momenta
$(x,{\bf k}_\perp; y,{\bf l}_\perp)$, the integral of ${\cal K}$
over $y$ and ${\bf l}_\perp$ in the nonvalence contribution, which we
define as $G_\pi\equiv\int[dy][d^2{\bf l}_\perp]
{\cal K}(x,{\bf k}_\perp;y,{\bf l}_\perp)\chi_{(2\to 2)}(y,{\bf l}_\perp)$,
depends only on $x$ and ${\bf k}_\perp$.
Approximating $G_\pi$ as a constant
has been tested in our previous exclusive semileptonic decay
processes~\cite{JC}
and proved to be a good approxiamtion at least for small momentum transfer
region. The validity of this
approximation can be checked by
fixing the constant $G_\pi$ by the sum rule expressed
in terms of ${\cal F}^{val}_\pi$ and ${\cal F}^{nv}_\pi$ as
\be\label{2sum}
F_\pi(t)=\int^1_\xi dx\; {\cal F}^{val}_\pi(\xi, x,t)
+ \int^\xi_0 dx\; {\cal F}^{nv}_\pi(\xi, x,t),
\ee
for given $-t$. We note that Eq.~(\ref{2sum}) is used as a constraint
on the frame-independence of our model.

\begin{figure}
\centerline{\psfig{figure=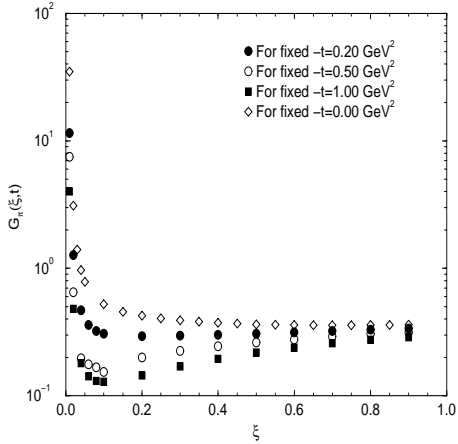,height=6.8cm,width=6.8cm}}
\vspace{-1.3cm}
\caption{The $\xi$-dependence of $G_\pi$ for different
momentum transfers $-t=0$ (diamond), 0.2 (black circle),
0.5(white circle), and 1 (black square) [GeV$^2$],
respectively. \label{Gfactor}}
\end{figure}

In Fig.~\ref{Gfactor}, we show the $\xi$-dependence of
$G_\pi$ for different $-t$-values, i.e. $-t=0$ (diamond), 0.2 (black
circle), 0.5 (white circle), and  1.0 (black square) [GeV$^2$], respectively.
As one can see in Fig.~\ref{Gfactor}, $G_\pi$ shows approximately
constant behavior for $\xi>0.1$ at given small $-t$.
It is not surprising to see that $G_\pi$ becomes very large
as $\xi\to 0$, because ${\cal F}^{nv}_\pi$ has
the form of ${\cal F}^{nv}_\pi = G_\pi \times \int^\xi_0...$
and the integral vanishes while a small but nonzero
contribution persists in ${\cal F}^{nv}_\pi$.
However, this does not cause a significant error in our
$G_\pi$ constant approximation because the nonvalence contribution in the
very small $\xi$ region is highly suppressed.
Therefore, the results are consistent with an almost constant value for
$G_\pi$ at least for small $-t$.
In principle, we can obtain the SPDs in a frame-independent
way by using the true values of $G_\pi$ as shown in Fig.~\ref{Gfactor}
for given $(\xi,t)$.

In this talk, we reviewed the connection between QCD and CQM pictures
in the LF quantization with a comparison and contrast to the ordinary
equal-t approach. We then discussed the newer development of LFQM
phenomenology in exclusive processes. The SPD appears to be a
good testing ground of our new effective treatment in timelike exclusive
processes. We investigated the SPDs of the pion
for small momentum transfer ($-t\leq 1$ GeV$^2$) region in the
LF quark model.
Since the LF nonvalence contributions to the SPDs
of the pion are large
especially at small momentum transfer region as shown in Ref.~\cite{CJK},
it is very crucial to take them into account to guarantee the
frame-independence of the model.
Applying our effective treatment~\cite{JC},
we express ${\cal F}^{nv}_\pi$ in terms of ordinary
LF wave functions of a gauge boson and a hadron and
calculate this nonvalence contribution numerically.
The reliability of our constant approximation was checked by examining
the frame-independence of our numerical results
using the sum rule given by Eq.~(\ref{2sum}), i.e.
the exact results of ${\cal F}_\pi(\xi,x,t)$ and $F_\pi(t)$
obtained from the true values of
$G_\pi$ given by Fig.~\ref{Gfactor} were compared with those
obtained from our single averge value of $G_\pi=0.32$ for all
$(\xi, t)$. The numerical results of our constant $G_\pi$ prescription
have shown definite improvement (better than 90 $\%$ accuracy for
$\xi\lsim 0.9$) to restore the
frame-independence of our model (see Ref.~\cite{CJK} for details)
and seemed to be a quite reliable approximation.
However, we note that there is an obvious $t$-dependence for
$G_\pi$, which leaves a room for more improvement of our model.
Consideration of the kernel $\cal K$ in the gauge boson sector and
the more realistic gauge boson wavefunction is underway.
\begin{center}
{\large\bf Acknowledgements}
\end{center}
We thank to the organizers of this workshop.
This work was supported in part by the US DOE under grant
No. DE-FG02-96ER40947 and by the NSF grant PHY-00070888 and
INT-9906384. The North Carolina Supercomputing Center and
the National Energy Research Scientific Computer Center are also
acknowledged for the grant of Cray time.


\begin{thebibliography}{99}
\bibitem{SSJC} A.P. Szczepaniak, E.S. Swanson, C.-R. Ji and
S.R. Cotanch, \Journal{\PRL}{76}{2011}{1996}.

\bibitem{RSSJC} D.G. Robertson, A.P. Szczepaniak, E.S. Swanson,
C.-R. Ji and S.R. Cotanch, \Journal{\PRD}{59}{074019}{1999}.

\bibitem{GJC1} E.L. Gubankova, C.-R. Ji and
S.R. Cotanch, \Journal{\PRD}{62}{074001}{2000}.

\bibitem{BPP} S.J. Brodsky, H.-C. Pauli and S.S. Pinsky,
Phys. Rept. {\bf 301}, 299 (1998).

\bibitem{zero} H.-M. Choi, and C.-R. Ji, 
\Journal{\PRD}{58}{071901}{1998}.

\bibitem{GJC2} E.L. Gubankova, C.-R. Ji and
S.R. Cotanch, \Journal{\PRD}{62}{125012}{2000}.

\bibitem{CJ} H.-M. Choi and C.-R. Ji,
\Journal{\NPA}{618}{291}{1997}; \Journal{\PRD}{56}{6010}{1997};
\Journal{\PLB}{460}{461}{1999}; \Journal{\PRD}{59}{074015}{1999}.

\bibitem{CKJ} See H.-M. Choi, L.S. Kisslinger and
C.-R. Ji in this Proceedings.

\bibitem{LB} G.P. Lepage and S.J. Brodsky, \Journal{\PRD}{22}{2157}{1980}.

\bibitem{Chung} P.L. Chung, F. Coester and W.N. Polyzou,
\Journal{\PLB}{205}{545}{1988};
W. Jaus, \Journal{\PRD}{44}{2815}{1991};
F. Cardarelli {\em et al.}, \Journal{\PLB}{332}{1}{1994}; 
\Journal{\PRD}{53}{6682}{1996}.

\bibitem{JC} C.-R. Ji and H.-M. Choi, \Journal{\PLB}{513}{330}{2001}.

\bibitem{CJK} H.-M. Choi, C.-R. Ji and L.S. Kisslinger,
\Journal{\PRD}{64}{093006}{2001}

\bibitem{Mu} D. M$\ddot{u}$ller, D. Robaschik, B. Geyer, F. M. Dittes,
J. Ho$\check{r}$ej$\check{s}$i, Fortsch. Phys. {\bf 42}, 101 (1994).

\bibitem{XJ1} X. Ji, \Journal{\PRL}{78}{610}{1997};
\Journal{\PRD}{55}{7114}{1997}.
\bibitem{Ra1} A. V. Radyushkin, \Journal{\PRD}{56}{5524}{1997}.
\end{thebibliography}
\end{document}